\documentclass[epj]{webofc}
\usepackage[varg]{txfonts}
\usepackage{epstopdf}
\woctitle{MESON2016 - the 14$^\textrm{th}$ International Workshop on Meson Production, Properties and Interaction}
\begin{document}
\selectlanguage{english}
\title{Calculations of kaonic nuclei based on chiral meson-baryon coupled channel interaction models}

\author{J. Hrtánková\inst{1,2}\fnsep\thanks{\email{hrtankova@ujf.cas.cz}} \and
        A. Ciepl\'{y}\inst{1} \and
        J. Mare\v{s}\inst{1} 
}

\institute{Nuclear Physics Institute, 250 68 \v{R}e\v{z}, Czech Republic
\and
           Czech Technical University in Prague, Faculty of Nuclear Sciences and Physical 
Engineering, \\ B\v{r}ehov\'{a} 7, 115 19 Prague 1, Czech Republic
          }

\abstract{
  We present our latest calculations of $K^-$-nuclear quasi-bound states using a self-consistent scheme for constructing $K^-$-nuclear potentials from various subthreshold chirally inspired $\bar{K}N$ scattering amplitudes. We consider in-medium versions of the scattering amplitudes taking into account Pauli blocking in the intermediate states. The resulting $K^-$ binding energies as well as the widths exhibit the same A dependence, however, the binding energies strongly depend on the model used.
}
\maketitle
\section{Introduction}
\label{intro}
\begin{figure}[t]
\begin{center}
\includegraphics[width=0.35\textwidth]{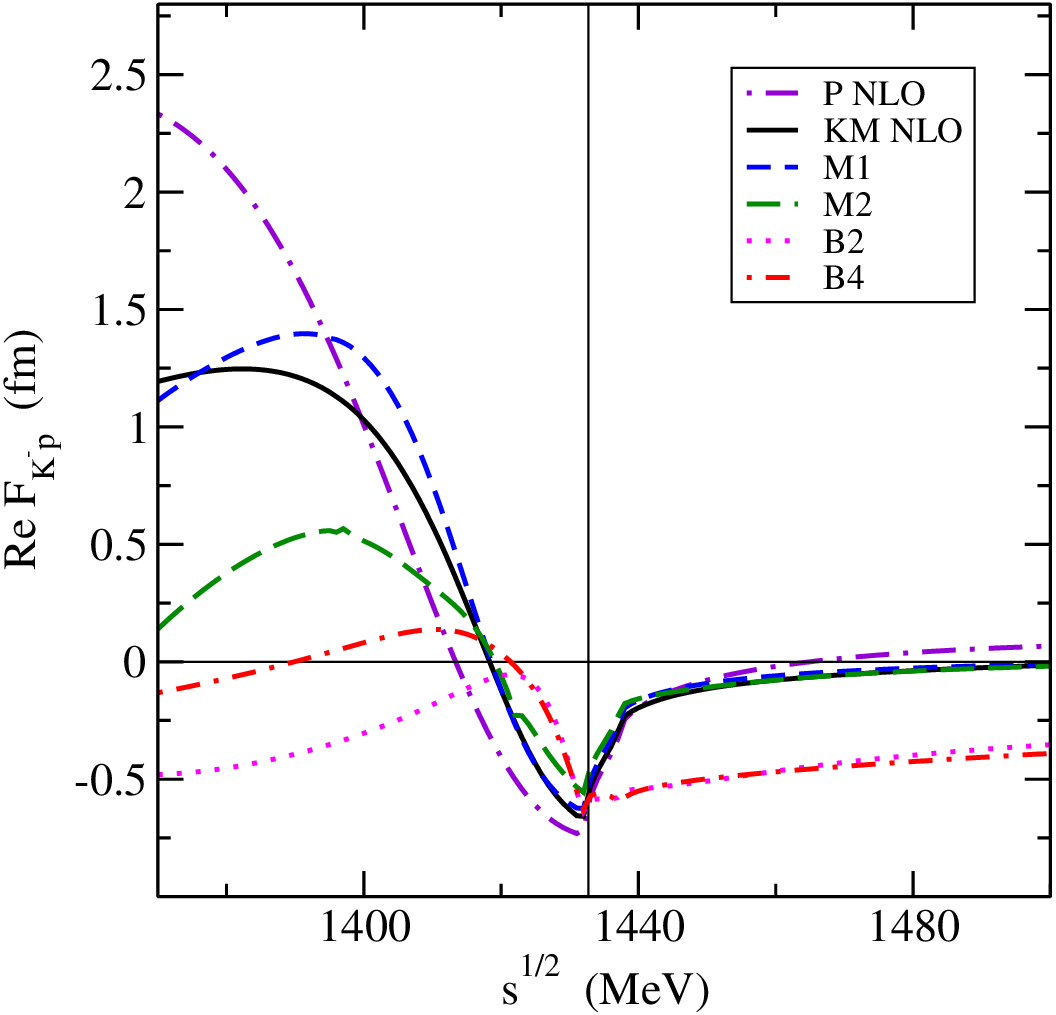} \hspace{10pt}
\includegraphics[width=0.35\textwidth]{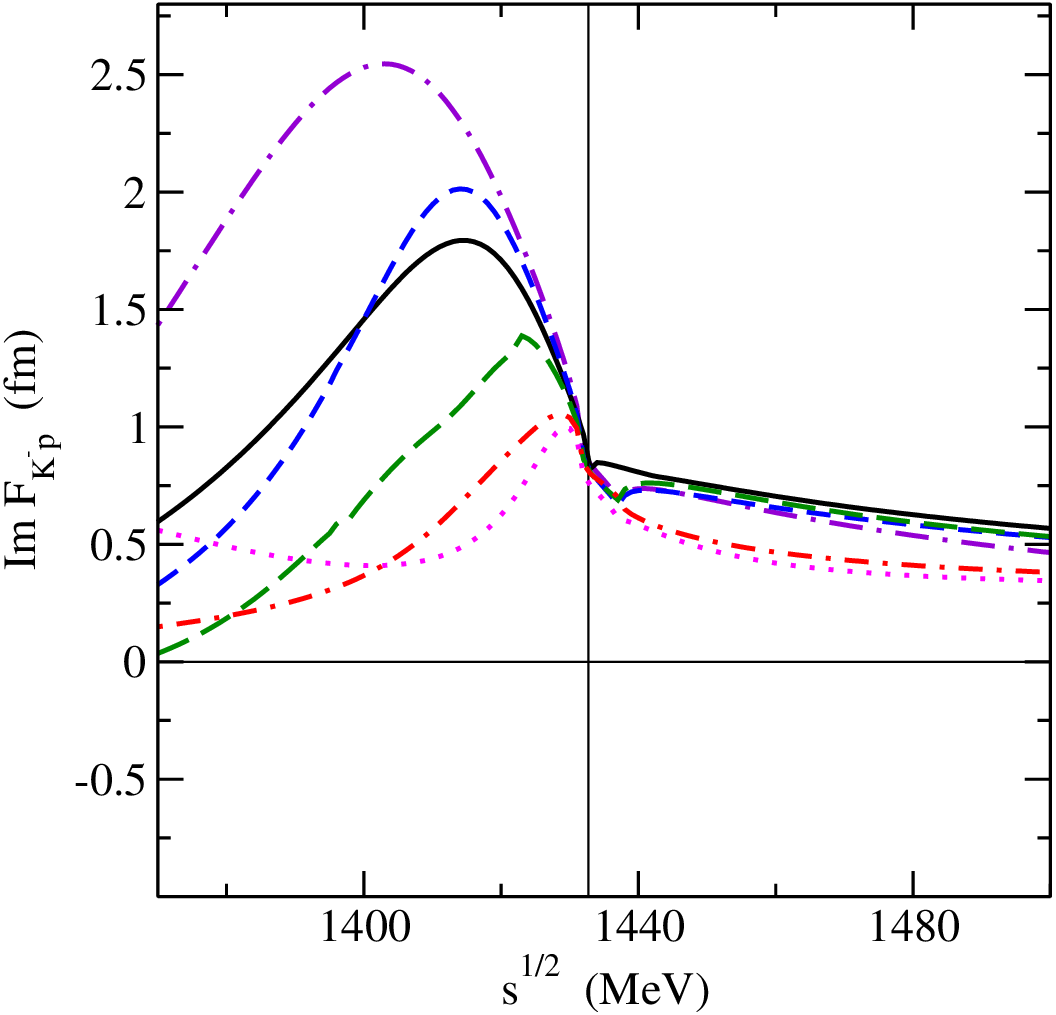} \hspace{50pt}
\includegraphics[width=0.35\textwidth]{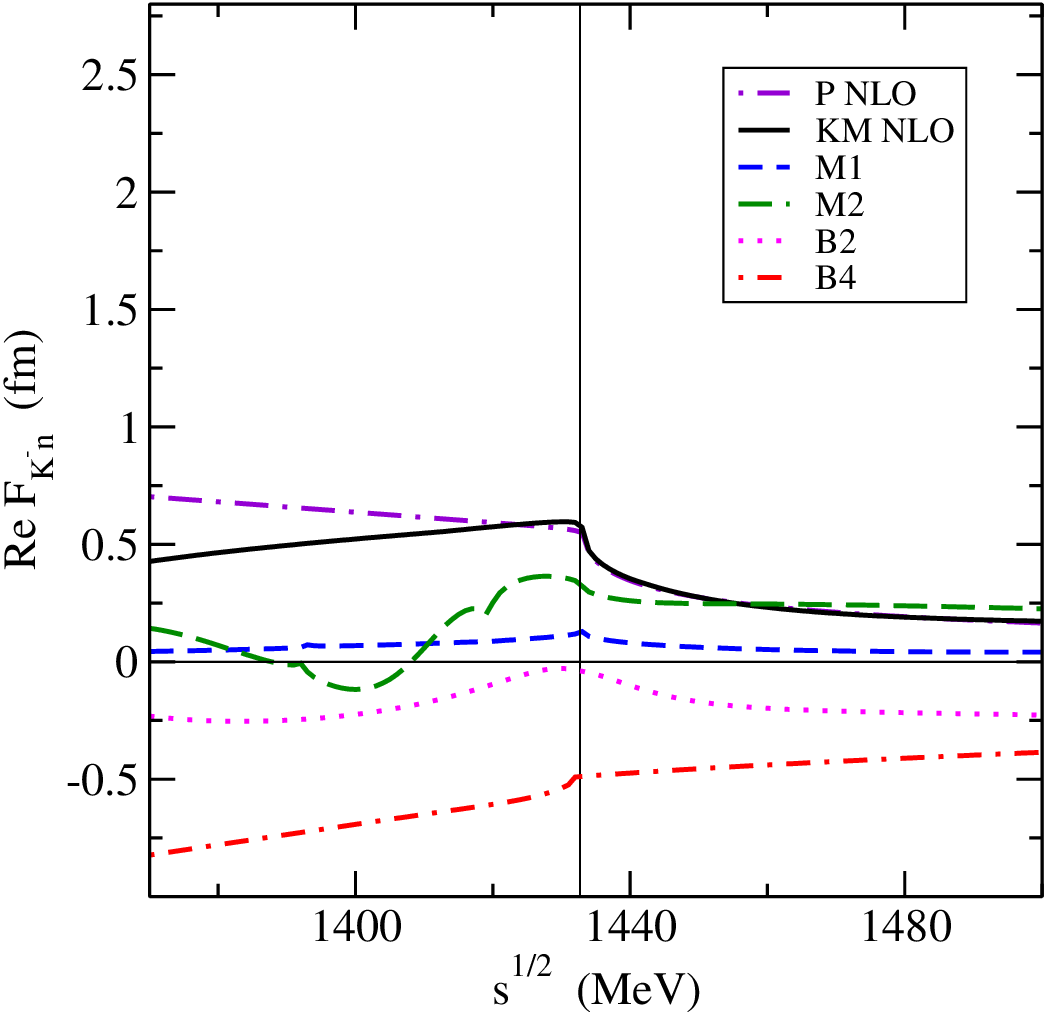} \hspace{10pt}
\includegraphics[width=0.35\textwidth]{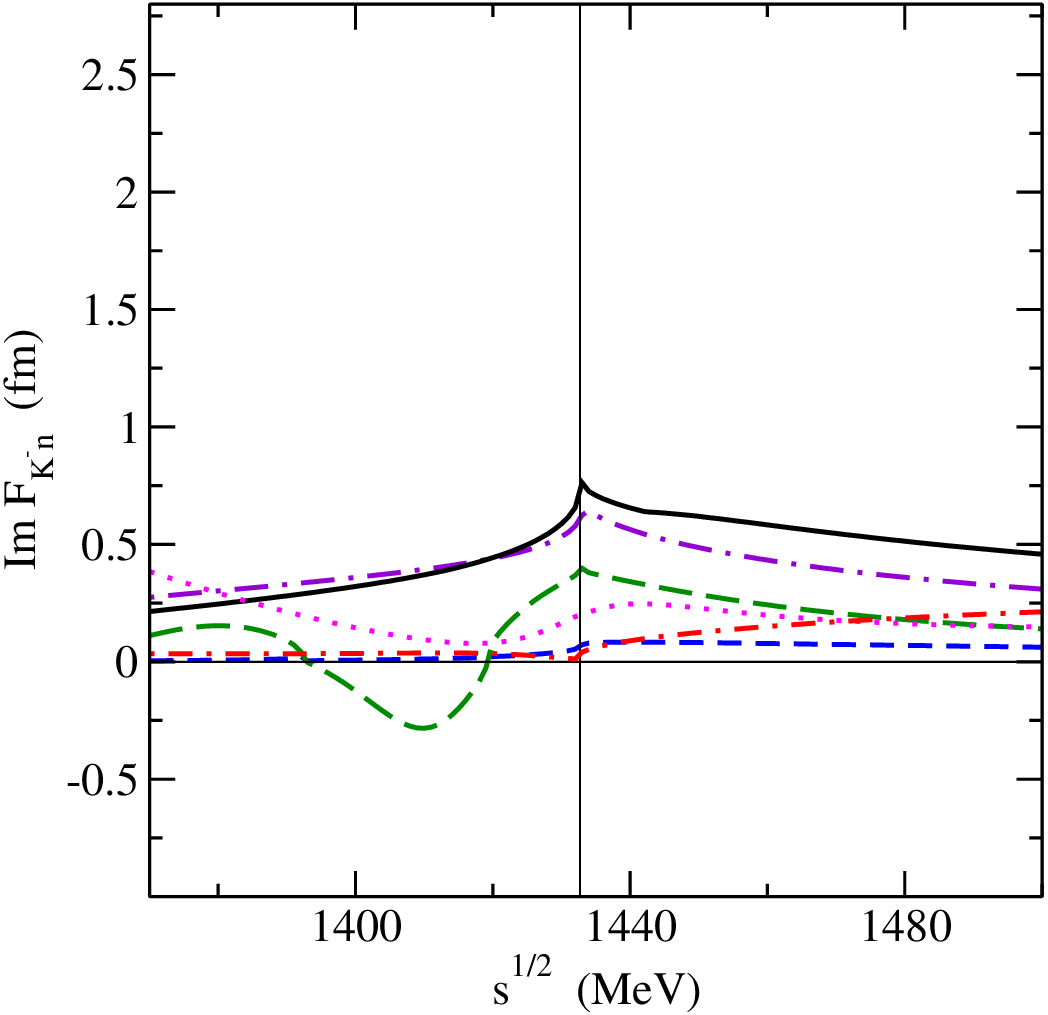}
\end{center}
\caption{Energy dependence of real (left) and imaginary (right) parts of free-space $K^-p$ (top) and $K^-n$ (bottom) amplitudes in considered models. }
\label{fig.:KpnFree}
\end{figure}

The aim of the present study is to compare the predictions for $K^-$-nuclear quasi-bound states calculated using different meson-baryon coupled channel interaction models: Prague (P NLO) \cite{pnlo}, Kyoto-Munich (KM NLO) \cite{kmnlo}, Murcia (M1 and M2) \cite{m}, and Bonn (B2 and B4) \cite{b}. They capture the physics of the $\Lambda$(1405) and reproduce low energy $K^-N$ observables, including the $1s$ level shift and width in the $K^-$ hydrogen atom from the SIDDHARTA experiment \cite{sidhharta}. However, the corresponding scattering amplitudes differ considerably below threshold, thus in the energy region relevant for $K^-$-nuclear bound-state calculations as shown in Figure~\ref{fig.:KpnFree}.  

\section{Model}
\label{sec-1}
\begin{figure}[t]
\begin{center}
\includegraphics[width=0.35\textwidth]{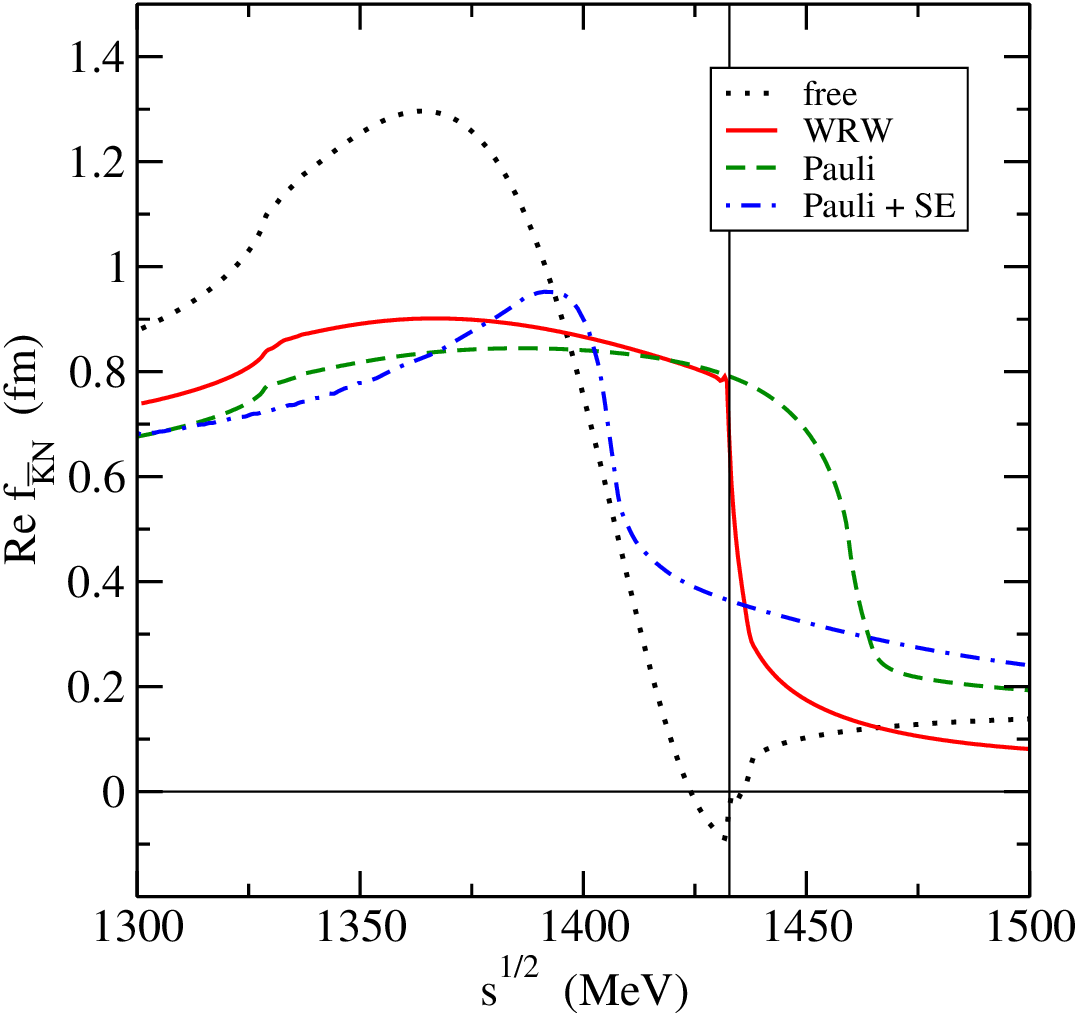} \hspace{10pt}
\includegraphics[width=0.35\textwidth]{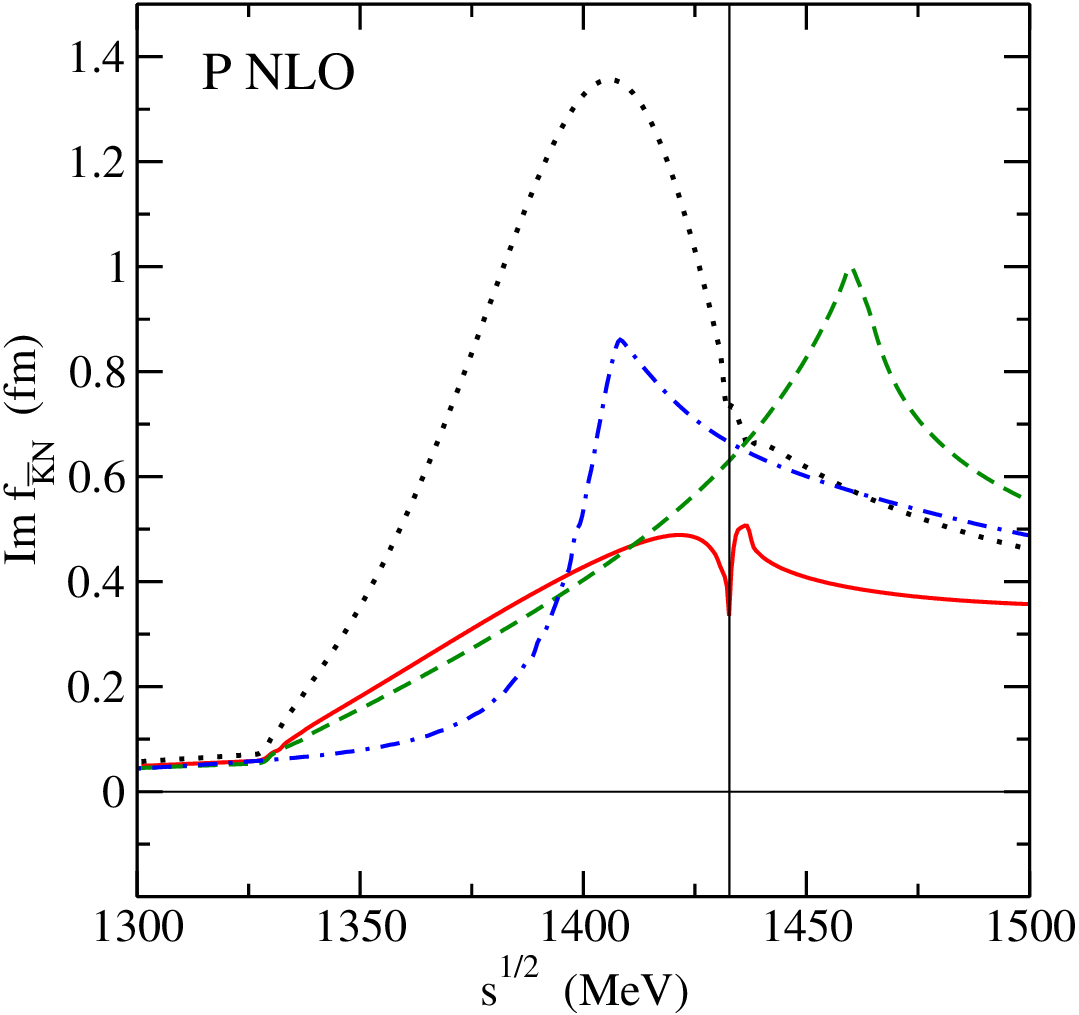}
\end{center}
\caption{Energy dependence of free-space (dotted line) $f_{\bar{K}N}=\frac{1}{2}(F_{K^- p}+F_{K^- n})$ amplitude compared with WRW modified amplitude (solid line), Pauli (dashed line), and Pauli + SE (dot-dashed line) modified amplitude for $\rho_0=0.17$~fm$^{-3}$ in the P NLO model. }
\label{fig.:inmedamp}
\end{figure}

The binding energies $B_{K^-}$ and widths $\Gamma_{K^-}$ of $K^-$-nuclear quasi-bound states are obtained by solving the Klein-Gordon equation
\begin{equation}
 \left[ \omega_{K^-}^2 +\vec{\nabla}^2  -m_{K^-}^2 -\Pi_{K^-}(\vec{p}_{K^-},\omega_{K^-},\rho) \right]\phi_{K^-} = 0~,
\end{equation}
where $\omega_{K^-} = m_{K^-} - B_{K^-} -{\rm i}\Gamma_{K^-}/2 -V_C= \tilde{\omega}_{K^-} - V_C$, $m_{K^-}$ is the $K^-$ mass, $V_C$ is the Coulomb potential, and $\vec{p}_{K^-}$ represents the kaon momentum. The self-energy operator $\Pi_{K^-}$ is constructed in a $t\rho$ form:
\begin{equation}
\Pi_{K^-} = 2\text{Re}( \tilde{\omega}_{K^-})V_{K^-}=-4\pi \frac{\sqrt{s}}{m_N}\left(F_0\frac{1}{2}\rho_p + F_1\left(\frac{1}{2}\rho_p+\rho_n\right)\right)~,
\end{equation}
where $F_0$ and $F_1$ denote the isospin 1 and 0 S-wave in-medium amplitudes, respectively, $m_N$ is the nucleon mass and $\sqrt{s}$ is the Mandelstam variable. The proton and neutron density distributions $\rho_p$ and $\rho_n$ are obtained within a relativistic mean-field model.
The in-medium amplitudes $F_0$ and $F_1$ are obtained from the free-space amplitudes by applying the multiple scattering approach (WRW)~\cite{wrw} which accounts for Pauli correlations:
\begin{equation}
F_{1}=\frac{F_{K^-n}(\sqrt{s})}{1+\frac{1}{4}\xi_k \frac{\sqrt{s}}{m_N} F_{K^-n}(\sqrt{s}) \rho}~, \quad F_{0}=\frac{[2F_{K^-p}(\sqrt{s})-F_{K^-n}(\sqrt{s})]}{1+\frac{1}{4}\xi_k \frac{\sqrt{s}}{m_N}[2F_{K^-p}(\sqrt{s}) - F_{K^-n}(\sqrt{s})] \rho}~,
\end{equation}
where $\xi_k$ is adopted from Ref.~\cite{wrw}.
In the P NLO model \cite{pnlo}, the integration in the underlying Green's function is limited to a certain domain due to the Pauli principle (Pauli) and the in-medium hadron self-energies (Pauli+SE) are considered as well. Figure~\ref{fig.:inmedamp} illustrates that WRW and Pauli approaches yield similar in-medium amplitudes in the subthreshold energy region relevant to our calculations.

The available energy $\sqrt{s}$ in the laboratory frame acquires the form \cite{cfggmPLB} (taking into account non-negligible contribution from particle momenta)
\begin{equation} \label{Eq.:K}
 \sqrt{s}= m_N + m_{K^-} - B_N- \xi_N B_{K^-} + \xi_{K^-} {\rm Re}{\cal V}_{K^-}(r) - \xi_N T_N\left(\frac{\rho}{\rho_0}\right)^{2/3}~,
\end{equation}
where $B_N$ is the average binding energy per nucleon, $\xi_{N(K^-)}={m_{N(K^-)}}/(m_N+m_{K^-})$, $T_N$ is the nucleon kinetic energy determined from the Fermi Gas model, and ${\cal V}_{K^-} = V_{K^-} + V_{\rm C}$.

 \begin{figure}[b]
\begin{center}
\includegraphics[width=0.35\textwidth]{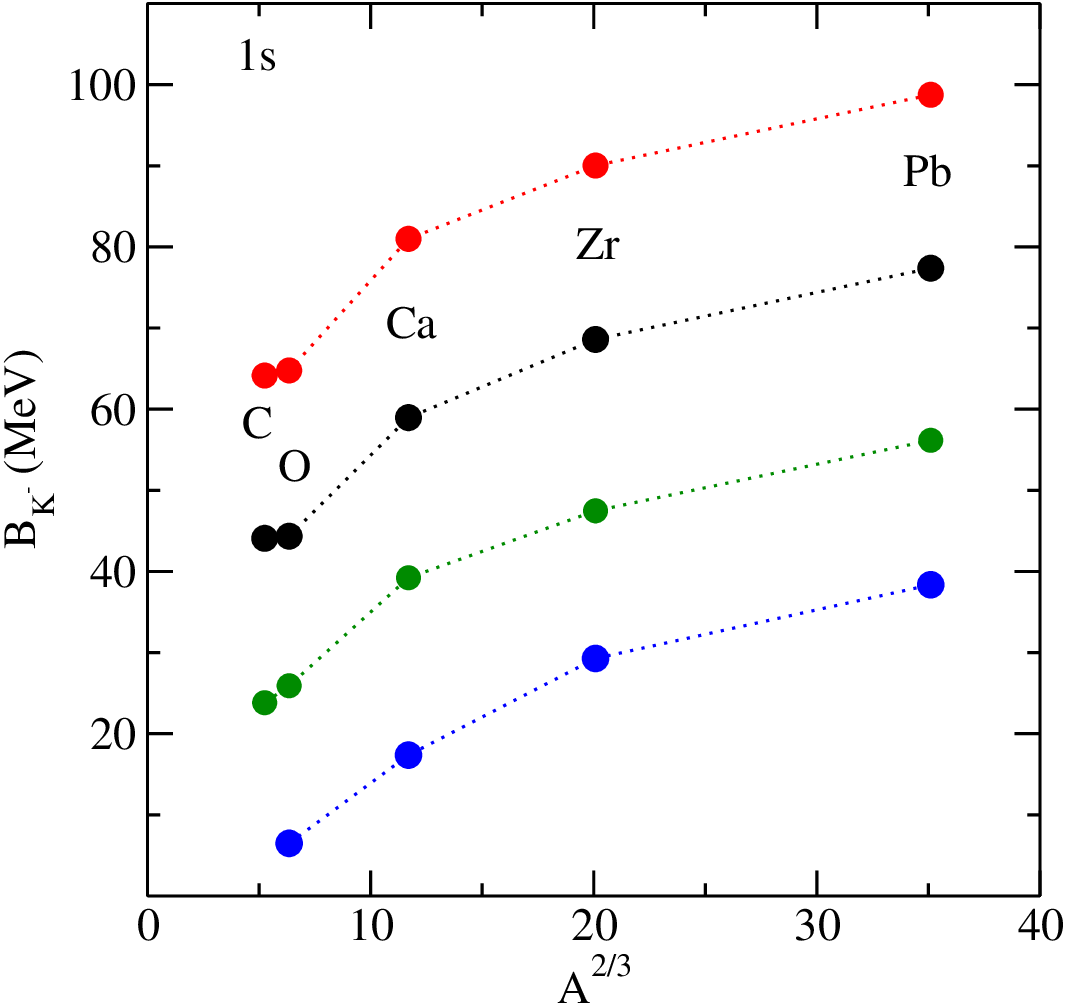} \hspace{10pt}
\includegraphics[width=0.35\textwidth]{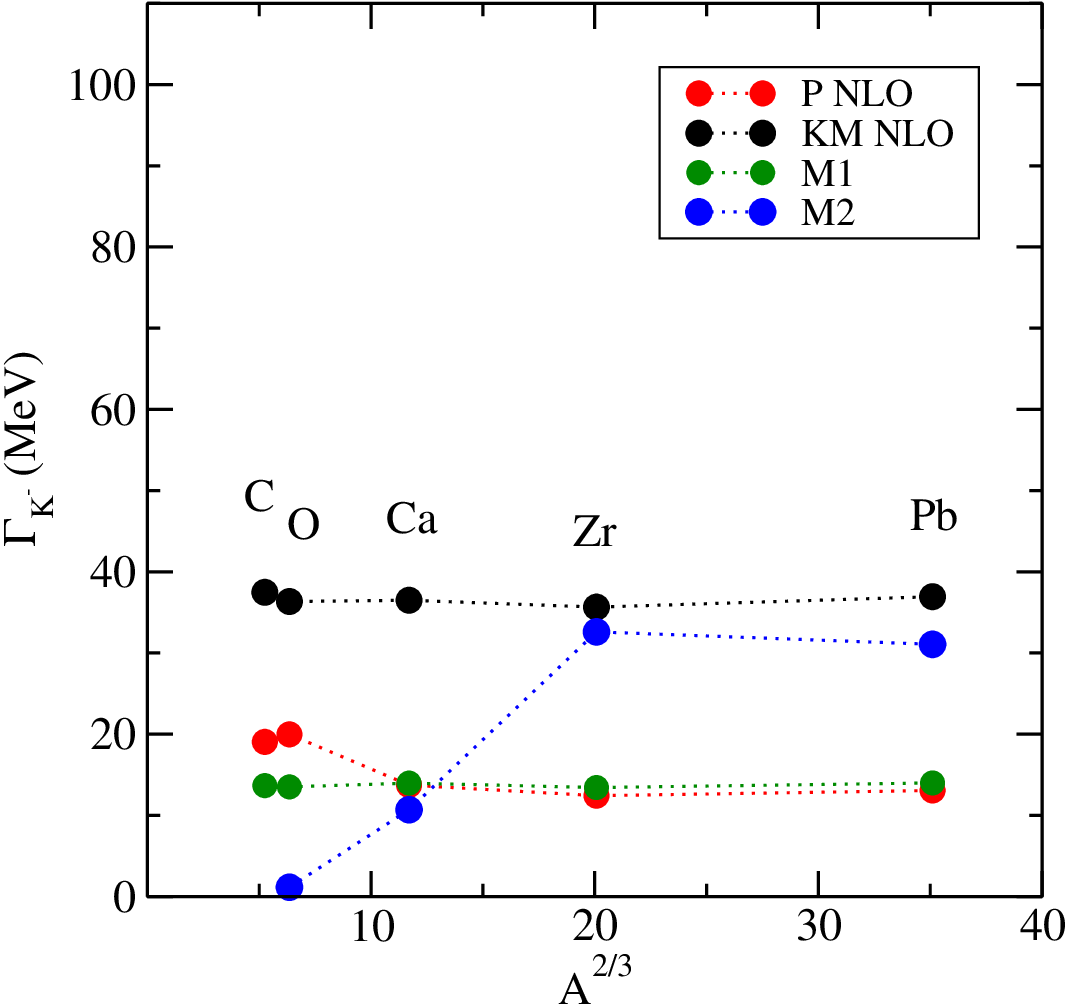}
\end{center}
\caption{1s $K^-$ binding energies (left) and corresponding widths (right) in various nuclei calculated self-consistently in the P NLO, KM NLO, M1, and M2 models. $K^- NN \rightarrow YN$ ($Y= \Lambda, \Sigma$) decay modes are not considered.}
\label{BkGk}
\end{figure}

\section{Results}
\label{sec-2}
With the formalism introduced above we performed self-consistent calculations of $K^-$ quasi-bound states in nuclei across the periodic table. In Figure~\ref{BkGk}, we present the $1s$ $K^-$ binding energies $B_{K^-}$ and corresponding widths $\Gamma_{K^-}$ as a function of the mass number calculated for different baryon-meson interaction models. The $K^-$ binding energies exhibit considerable model dependence, nevertheless their  A dependence is very similar in all models considered. The $K^-$ widths show rather weak A dependence except for the M2 model. The KM NLO model predicts widths twice as large as the P NLO and M1 models. Within the Bonn models B2 and B4, we did not succeed to obtain any bound states since the real parts of the $K^-N$ amplitudes are weakly attractive or even repulsive below threshold (see Figure~\ref{fig.:KpnFree}). 
\begin{figure}[t]
\begin{center}
\includegraphics[width=0.35\textwidth]{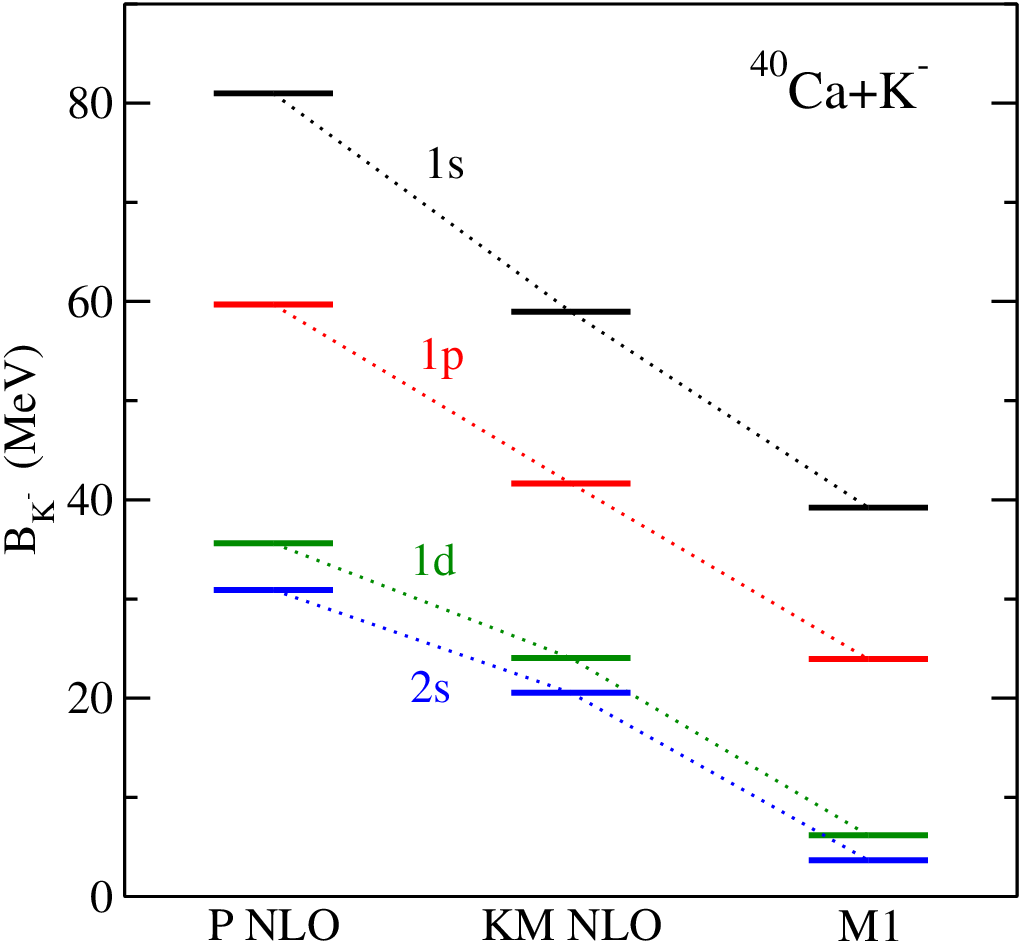} \hspace{10pt}
\includegraphics[width=0.35\textwidth]{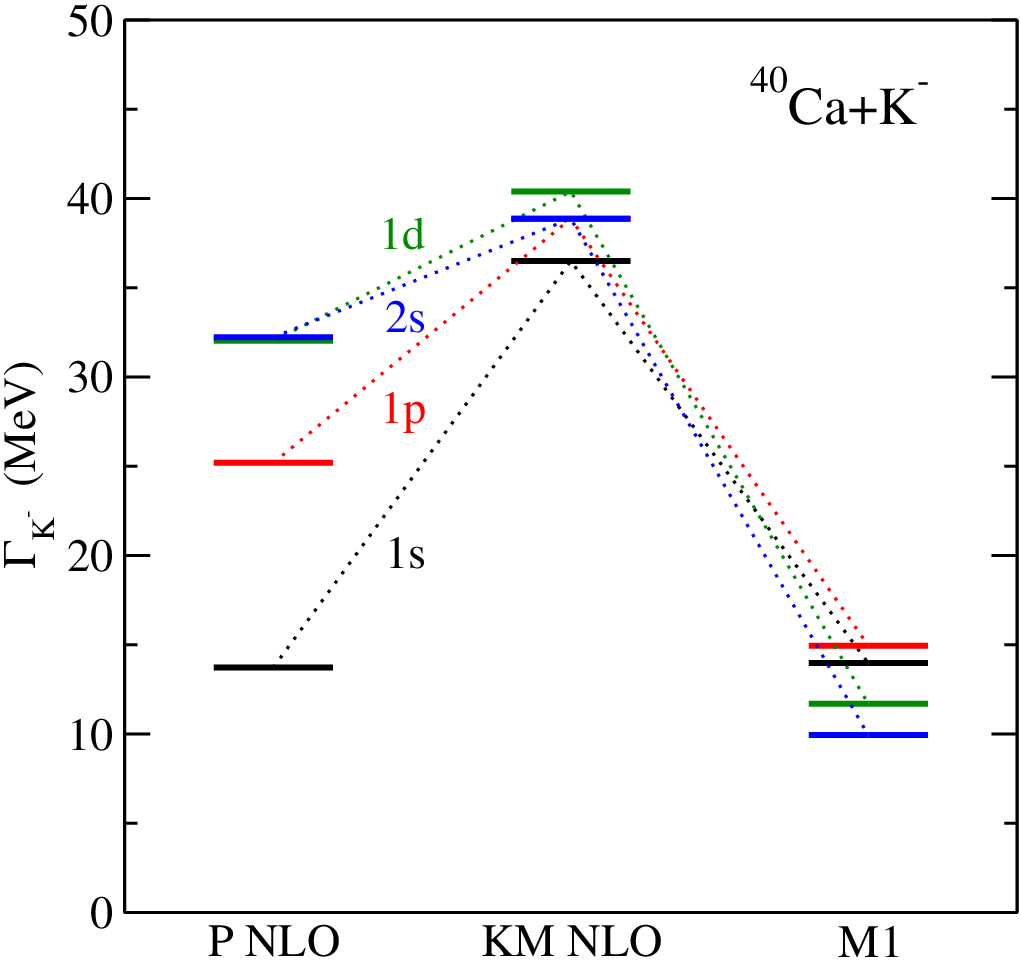}
\end{center}
\caption{$K^-$ binding energies (left) and widths (right) in $s, p$ and $d$ levels in $^{40}$Ca calculated self-consistently in the P NLO, KM NLO, and M1 models. $K^- NN \rightarrow YN$ decay modes are not considered.}
\label{BkGkCa}
\end{figure}
In Figure~\ref{BkGkCa}, the $K^-$ spectrum in $^{40}$Ca calculated for various models is shown. Again, the $K^-$ binding energies strongly depend on the model used. In the P NLO and KM NLO models, the $K^-$ in the $1s$ state has the smallest $K^- N \rightarrow \pi Y$ conversion width due to the considerable energy shift towards the $\pi \Sigma$ threshold. The $K^-$ widths of excited states grow as the $\sqrt{s}$ moves farther from the $\pi \Sigma$ threshold. However, in the M1 model the widths follow the opposite trend. It is due to the fact that $\sqrt{s}$ is much closer to the $\bar{K}N$ threshold where the imaginary part of the $K^-p$ amplitude starts to decrease (see Figure~\ref{fig.:KpnFree}).
 
It is to be noted that the core polarization caused by $K^-$ was not taken into account in the present calculations. Previous studies of kaonic nuclei \cite{gmNPA} found this effect quite mild, adding up to 5~MeV to $K^-$ binding energies. Moreover, the annihilation of $K^-$ on 2 nucleons \cite{gmNPA} should be considered as well, since it increases the width by almost 50~MeV and, consequently, even the $1s$ state $K^-$ widths become comparable with the binding energies. We intend to include both effects mentioned above in the upcoming calculations.
\\

\begin{acknowledgement}
We wish to thank A. Gal and E. Friedman for fruitful discussion, and M. May for providing us with the free $\bar{K}N$ scattering amplitudes.
This work was supported by the GACR Grant No. P203/15/04301S. J. Hrt\'{a}nkov\'{a} acknowledges financial support from CTU-SGS Grant No. SGS16/243/OHK4/3T/14. 
\end{acknowledgement}
%
%
%

\end{document}